# Scaling in Plateau-to-Plateau Transition: A Direct Connection of Quantum Hall Systems with Anderson Localization Model

Wanli Li,[1,*] C. L. Vicente,[2] J. S. Xia,[2] W. Pan,[3] D. C. Tsui,[1] L. N. Pfeiffer,[4] and K. W. West[4]

[1]Princeton University, Princeton, NJ 08544, USA
[2]National High Magnetic Field Laboratory (NHMFL) and University of Florida, Gainesville, FL 32611, USA
[3]Sandia National Laboratories, Albuquerque, NM 87185, USA
[4]Bell Laboratories, Lucent Technologies, Murray Hill, NJ 07974, USA

The quantum Hall plateau transition was studied at temperatures down to 1 mK in a random alloy disordered high mobility two-dimensional electron gas. A perfect power-law scaling with $\kappa$=0.42 was observed from 1.2K down to 12mK. This perfect scaling terminates sharply at a saturation temperature of $T_s$~10mK. The saturation is identified as a finite-size effect when the quantum phase coherence length ($L_\phi \propto T^{-p/2}$) reaches the sample size (W) of millimeter scale. From a size dependent study, $T_s \propto W^{-1}$ was observed and *p*=2 was obtained. The exponent of the localization length, determined directly from the measured $\kappa$ and *p*, is $\nu$=2.38, and the dynamic critical exponent *z* = 1.

Fifty years ago, P.W. Anderson [1] showed that in randomly disordered three-dimensional electronic systems a quantum particle can undergo a phase transition from a metallic state to an insulating state. Ever since, extensive studies on this transition have vastly expanded and deepened our understanding of electron-disorder interaction in condensed matter physics [2]. In the late 70s', in another seminal work on the two-dimensional electron system (2DES), Abrahams et al [3] argued that under a single particle picture no extended states can survive in the thermodynamic limit. Thus, at T=0 all the states are localized. In 1994, however, the surprising observation of an apparent metal-to-insulator transition (MIT) in high quality silicon metal-oxide-semiconductor field-effect transistors [4] raised a serious challenge to this well established 2D scaling model. 15 years later, whether this apparent MIT represents a true quantum phase transition induced by a strong electron-electron interaction is yet to be settled.

On the other hand, the introduction of a magnetic field perpendicular to the 2DES is known to break the continuous 2DES density of states, create discreet Landau levels, and bring in delocalized states in the center of these Landau levels. All this, in fact, gives rise to the integer quantum Hall (QH) effect [5]. In the QH regime, the Hall plateaus represent separate energy regions of localized states, while between two adjacent plateaus there is a critical field $B_c$ associated with one energy level of extended states [6, 7]. The plateau-to-plateau transition (PPT) is therefore a localization-delocalization transition (LDT). Following the idea of quantum phase transitions, a power law divergence of the localization length $\zeta \propto |B-B_c|^{-\nu}$ applies when the critical field, $B_c$, is being approached from both sides [8-11]. The localization length exponent $\nu$ characterizes the critical behavior of the transition at zero temperature and has been the focus of both theoretical and experimental studies. It is now generally believed that $\nu$ = 2.4 [9,10,12]. At finite temperatures, this power law of $\zeta \propto |B-B_c|^{-\nu}$ is translated by finite size scaling theory into a temperature scaling form $(dR_{xy}/dB)|_{Bc} \propto T^{-\kappa}$, where $\kappa = p/2\nu$ and $p$ is the temperature exponent of the inelastic scattering length or quantum coherence length of the 2DES [8-10], $L_\phi \propto T^{-p/2}$.

In his original paper [1], Anderson modeled a disorder system as random lattices. While in the QH systems, the nature of disorder can vary vastly from one material system to another. It has been anticipated that, as an Anderson LDT of disordered 2DES [1], the nature of the disorder should play an important role in PPT. In the first experiment by Wei et al [13] on PPT in the 2DES realized in InGaAs/InP heterostrcutures, where the disorder is dominated by short-range alloy potential fluctuations and thus its range much smaller than the quantum de-phasing length, a universal scaling exponent $\kappa$=0.42 was obtained in a temperature range from 0.1K to 4.2K. In the studies of different experimental systems, such as the conventional *GaAs/Al$_{0.32}$Ga$_{0.68}$As* heterostructures, the disorder is mostly from the remote ionized impurities and the screened Coulomb potential has a correlation length on the order of micrometers and amplitude of meV in the plane of the 2DES [14]. In these systems, $\kappa$ was found to be non-universal and varied from 0.16 to 0.81. Because of this discrepancy, the universality of PPT was called into question [15]. On the other hand, it has long been suspected that this discrepancy could be due to the nature of disorder in different material systems, and that the long range disorder in the later studied systems may be responsible for the obtained non-universal values. Indeed, in these systems, the electron transport is better described by a percolation picture through the so-called saddle points than by Anderson localization. Consequently, the scaling regime of PPT has yet to be reached and the obtained non-universal $\kappa$ value at most represents a cross-over behavior.

A recent work focusing on alloy disorder in *Al$_x$Ga$_{1-x}$As/Al$_{0.32}$Ga$_{0.68}$As* heterostructure [16,17] has brought new insight on this problem. In comparison with the ionized impurity disorder, the alloy disorder in the 2DES (residing in *Al$_x$Ga$_{1-x}$As*) is short in range (only over a distance of *GaAs* lattice constant, 0.56nm) and strong in amplitude (with the amplitude of alloy potential fluctuation ~1.13eV [16]), when the value of *x* is small and falls in the range of 0.65<*x*<1.6%. In other words, the disorder in this sample structure is of the same type of disorder discussed in [1], *thus providing an opportunity to directly connect the Anderson localization theory with real experimental systems*. Indeed, the universality of PPT is restored with $\kappa$=0.42±0.01 [17].

In this letter, we carry out an experimental investigation of the QH plateau-to-plateau transition in one such alloy disorder dominated sample down to a new low temperature regime, to 1mK, in a nuclear demagnetization refrigerator. A perfect temperature scaling, $(dR_{xy}/dB)|_{Bc} \propto T^{-0.42}$, is observed through two full decades of temperature from 1.2K down to 12mK. Surprisingly, a sharp saturation of $(dR_{xy}/dB)|_{Bc}$ occurs below 10mK. By systematically examining a number of different size specimens, the saturation is identified to be a finite-size effect when the quantum phase coherence length reaches *the sample size of millimeters* at ultra-low temperatures. This observation allows us to determine the temperature exponent of the inelastic scattering length ($p$ = 2) in our samples, and a direct measurement of the exponent of localization length ($\nu$ = 2.38).

The sample is a modulation doped *$Al_xGa_{1-x}As/Al_{0.32}Ga_{0.68}As$* heterostructure with $x$=0.85%. The 2DES density and mobility of electrons are $n$ = 1.2 x $10^{11}$ $cm^{-2}$ and $\mu$ = 8.9 x $10^5$ $cm^2$/Vs. Several rectangle shaped specimens with the ratio of length/width = 4.5:2.5 were cut from the same wafer, and the largest one is of size 4.5mm x 2.5mm. Our ultra-low temperature experiment was carried out in a nuclear demagnetization /dilution refrigerator with a base bath temperature ($T_b$) below 1mK. The same measurement setup as in Ref. [18] was employed. Standard lock-in technique was used to measure the longitudinal magneto-resistance $R_{xx}$ and the Hall resistance $R_{xy}$ with a current excitation of 1nA and frequency of 5.7Hz.

We concentrate on the Hall resistance $R_{xy}$ in this experiment. Data from $R_{xx}$ are checked at a few temperatures, and always consistent with the $R_{xy}$ measurement [17]. Figure 1 shows the Hall resistance in the largest specimen around the transition from the plateau of filling factor $\nu$=4 to the plateau of filling factor $\nu$=3 (4-3 transition) in a large temperature range. All the curves cross at one point, which labels the critical magnetic field $B_c$=1.4T.

The values of *$(dR_{xy}/dB)|_{Bc}$* was calculated at different temperatures, and plotted vs. *T* on a log-log scale in Figure 2 (a). The data presented here were taken in three different cryostats and they fall on top of each other where they overlap in temperature. In the temperature range from 1.2K down to 12mK, a perfect power law scaling *$(dR_{xy}/dB)|_{Bc}$* $\propto$ $T^{-\kappa}$ with $\kappa$=0.42±0.01 is observed. We emphasize that this is for the first time that the power law critical behavior of QH localization-delocalization transition is observed in two full decades of temperature. *This shows unequivocally that in an Anderson disordered 2DES the scaling behavior indeed prevails.*

As we lower the temperature below 10mK, *$(dR_{xy}/dB)|_{Bc}$* is observed to saturate sharply, instead of diverging. The saturation is shown in Figure 2 (b). We note here that a similar saturation effect has been observed before in mesoscopic samples of size ranging from 10μm to 64μm and the saturation was interpreted to be due to finite size of the sample [19]. To investigate this finite sample size effect in our measurements, we have fabricated rectangle-shaped specimens of various sizes to study the saturation of the temperature scaling. The width of these specimens ranges from 500μm down to 100μm, with the length-to-width ratio being kept to 4.5:2.5. In these samples we have observed the same temperature scaling $(dR_{xy}/dB)|_{Bc} \propto T^{-0.42}$, though for samples of different sizes the scaling terminates at different temperatures. In Fig. 3a, we show the saturation for two samples of 500μm and 100μm. It is clearly seen that the saturation temperature $T_s$ of $(dR_{xy}/dB)|_{Bc}$ is higher in smaller samples. In Fig. 3b, we plot the dependence of $T_s$ on the sample width $W$ for 5 samples. Within the experimental uncertainty, $T_s$ is found to be inversely proportional to $W$.

Before we discuss the significance of Fig. 3b, we need to rule out the possibility that the saturation is due to self-heating of the electrons by the applied excitation current or by external noise, i.e., the electron temperature $T_e$ can not be cooled below 10mK. First, we note that a previous experiment on a high mobility 2DES sample showed that external noise by itself did not heat the electrons beyond 4mK when the cryostat was at $T_b$=1mK [18]. To further investigate the internal heating due to the excitation current, we measured $R_{xy}$ with different excitations at the base bath temperature of $T_b$=1mK. The values of $(dR_{xy}/dB)|_{Bc}$ with different excitations are displayed in Fig. 4. It is clearly seen that $(dR_{xy}/dB)|_{Bc}$ is constant for excitations below 2nA, while the excitation current applied in our experiments is 1nA. We thus infer from the arguments above that the saturation of $(dR_{xy}/dB)|_{Bc}$ below 10mK cannot be an effect from electron heating.

Having ruled out self-heating as a cause, we show in the following that the termination of scaling at low temperatures is due to finite size of the sample. The temperature scaling form $(dR_{xy}/dB)|_{Bc} \propto T^{-\kappa}$ is obtained by the finite size scaling theory [9, 10]. In this theory, the transport properties are determined by the ratio between the localization length $\zeta \propto |B-B_c|^{-\nu}$ and the effective sample size which is the quantum phase coherence length $L_\phi$. However, as the temperature approaches zero, $L_\phi$ diverges following $L_\phi \propto T^{-p/2}$ and can be larger than the length (or width, whichever smaller) of the sample. Under this condition, the actual sample size (in our experiment the width of sample $W$) becomes a "hard" limit for $L_\phi$ and will terminate the temperature scaling.

The observed strong size-dependence of $T_s$ in Fig. 3b demonstrates that the saturation is indeed a finite-size effect. At discussed above, the saturation is reached when $L_\phi$ reaches the actual sample width $W$ at $T_s$. Thus, this $T_s \propto W^{-1}$ dependence implies that $L_\phi$ is inversely proportional to temperature. From this $1/T$ dependence, the temperature exponent of inelastic scattering length $p=2$ is obtained [20,21]. Now, with $p$ and $\kappa$ measured directly in our experiment, it is possible to determine the value of the exponent of the localization length $\nu=2.38$. This value is consistent with various numerical calculations of $\nu\sim2.4$. We note here that this is the first experiment in which universal values of both $\kappa$ and $\nu$ are determined in a high mobility 2DES, and that the dynamic frequency scaling exponent obtained from $\kappa = 1/\nu z$ is $z =1$, in good agreement with previous microwave conductivity measurements [20].

The millimeter length scale of $L_\phi$ at low temperatures is rather surprising, and a $L_\phi$ of this macroscopic length scale has never been reported. In the literature, $L_\phi$ is expected to be large only along the sample edge due to the suppression of electron-electron scattering in the QH edge channels [22]. In the region around PPT, physics of the bulk dominates, and our observation suggests that quantum phase coherence can be kept over a long distance in the bulk as well. In the following, we estimate the $L_\phi$ of the bulk 2DES at zero magnetic field. For clean 2DES at low temperatures, electron-electron scattering dominates the dephasing mechanism. Following the method in [23], we estimate the scattering rate $\tau^{-1}_{e-e}$ [23-25], and obtain the value of $L_\phi = (D\tau_{e-e})^{1/2}$, with $D$ being the electron diffusion coefficient. At temperature $T_s$=10mK, $L_\phi$ is estimated to be 1.4mm, which is of the same order of our sample size. This estimation suggests that the millimeter-size $L_\phi$ in the bulk is a result from the high sample quality (thus a large $D$) and the low temperature that significantly reduces the electron-electron scattering.

One elegant way to visualize the physics underlying the QH effect is the edge channel picture [26-28]. $L_\phi$ along the edge is generally very long due to the perfect reflection of the skipping orbits from the edge potential. In the PPT region, electrons from one edge channel can travel to the opposite-propagating edge channel on the other side via resonant tunneling [29,30], which smears out the sharpness of the plateau-to-plateau transition. The saturation of $(dR_{xy}/dB)|_{Bc}$ shows that the probability of inter-channel tunneling saturates below $T_s$ when the whole sample is phase coherent.

Although we anticipate that $L_\phi$ reaches the sample size at temperature $T_s$, we did not find any feature of the universal conductance fluctuation (UCF) on either $R_{xy}$ or $R_{xx}$. We suggest that the absence of UCF is a thermal averaging effect. The thermal length $L_T$ is given by $L_T = (\hbar D/k_B T)$, and is only about 20μm in our samples at $T_s$=10mK. Since $L_T$ is much smaller than the sample size $W$, the UCF is thermally averaged out even though the electrons are dynamically phase coherent all over the sample [32].

In conclusion, we have observed a perfect power law scaling of the localization-delocalization transition in an alloy disordered 2DES subjected to magnetic field. The power law scaling spans over two full decades of temperature from 1.2K down to 12mK before the saturation of $(dR_{xy}/dB)|_{Bc}$. The saturation at low temperatures is shown to be a finite-size effect, from which we can determine the temperature exponent of inelastic scattering independently and are able to obtain directly the value of the exponent of the location length. Moreover, the millimeter scale $L_\phi$ around the plateau-to-plateau transition in our samples at ultra-low temperatures is an example that quantum mechanics prevails in a macroscopic regime in semiconductor systems. We have therefore tested the physics of Anderson localization in 2DES in a millimeter length scale over a wide temperature range.

This work was supported by the NSF. We thank H.L. Stormer, X. Wan, and A. Pruisken for inspiring discussions. WP was supported by DOE/BES at Sandia, a multiprogram laboratory operated by Lockheed Martin for the DOE/NNSA under contract DE-AC04-94AL85000.  Part of the work was done at the NHMFL high B/T facilities.

* email address: sciwanli@gmail.com.

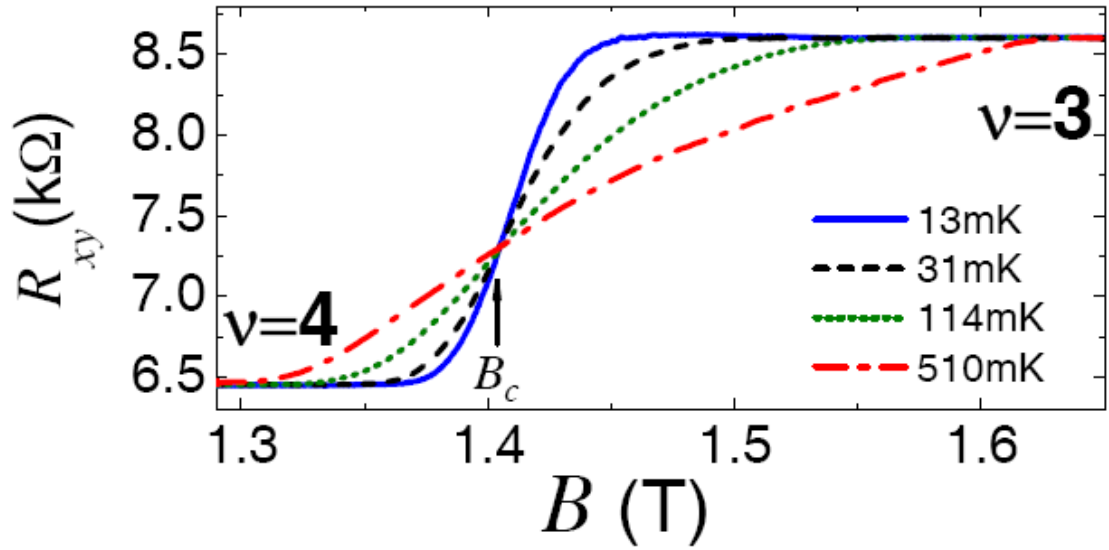

Fig. 1: Hall resistance around the 4-3 transition at different temperatures. A critical field of $B_c$=1.4T is observed.

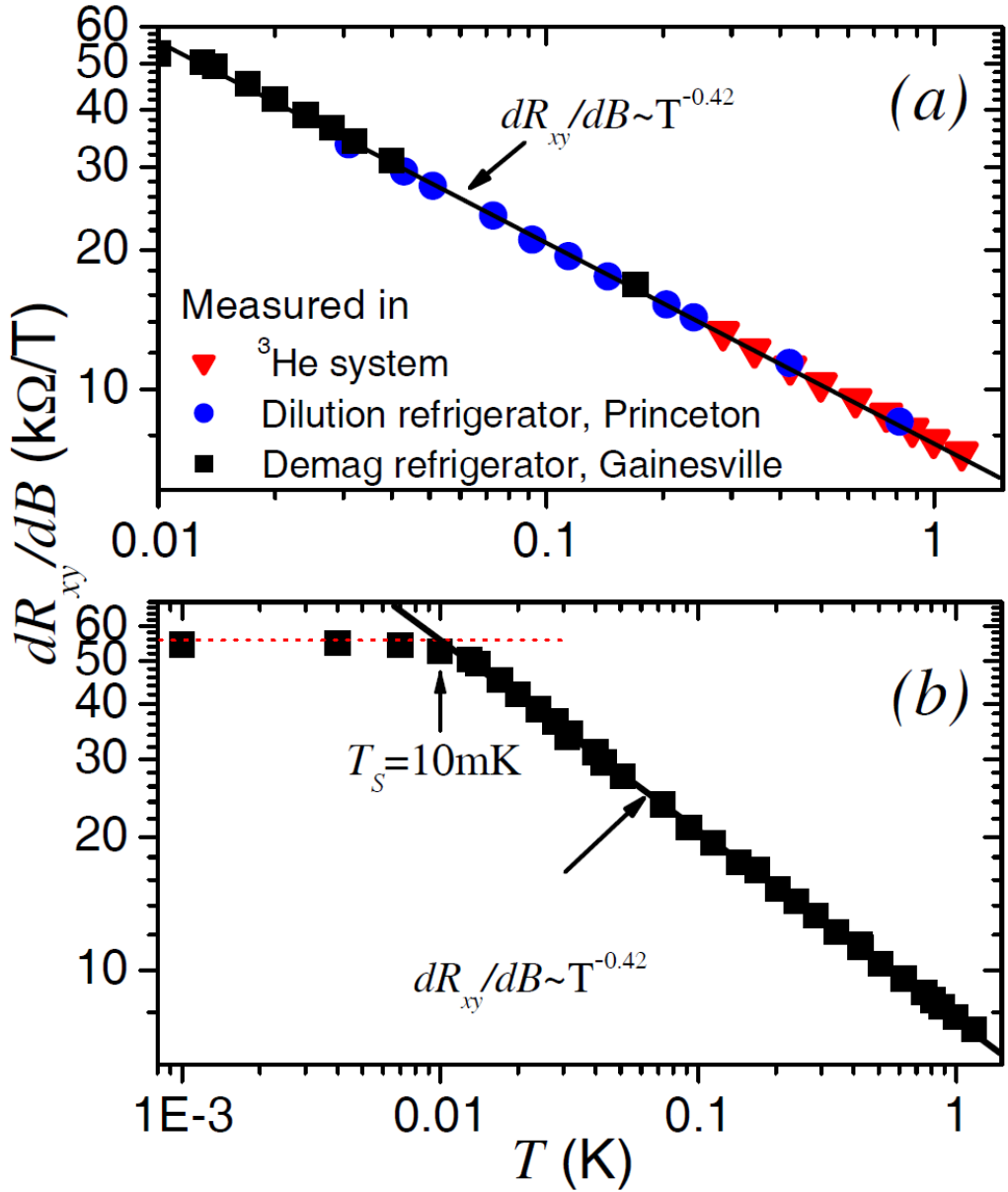

Fig.2: (a) Perfect temperature scaling $(dR_{xy}/dB)|_{Bc} \propto T^{-0.42}$ of the 4-3 transition over two decades of temperature between 1.2K and 12 mK. (b) Saturation of $(dR_{xy}/dB)|_{Bc}$ at low temperatures. The saturation temperature $T_s$=10mK is obtained from the cross point between extrapolations of the higher temperature data (black line) and the lower temperature saturated data (horizontal dotted line).

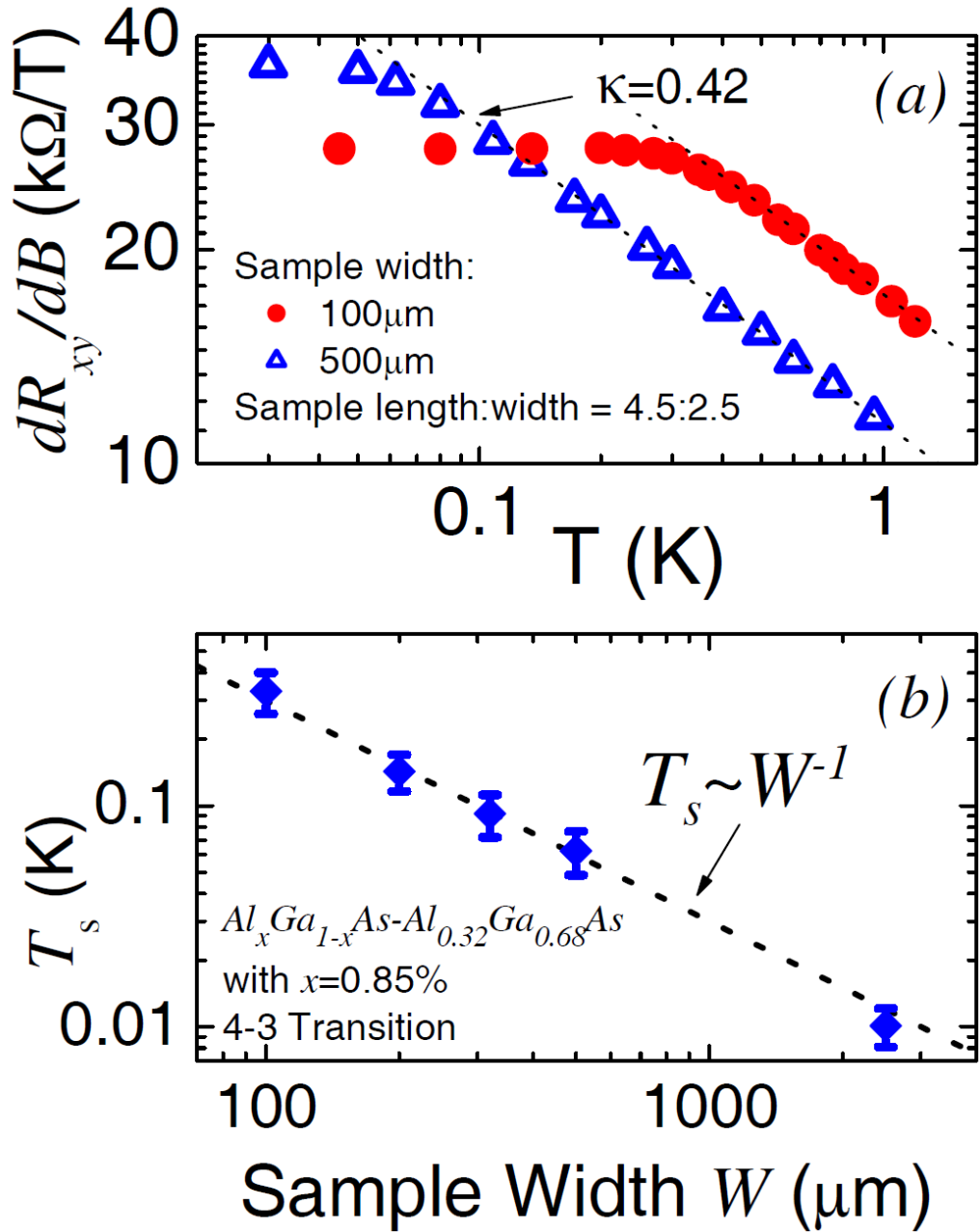

Fig.3: (a) $(dR_{xy}/dB)|_{Bc}$ vs. $T$ of the 4-3 transition for two samples of different size. The length-width of samples is kept to be 4.5:2.5. (b) The sample size dependence of the saturation temperature $T_s$ of $(dR_{xy}/dB)|_{Bc}$. The value of $T_s$ is inversely proportional to the sample width $W$ within the error.

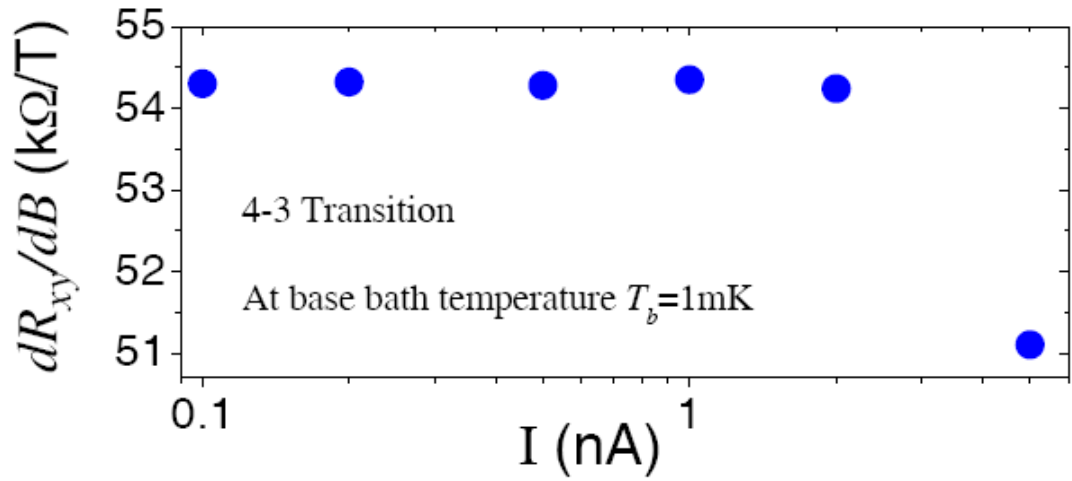

Fig.4: $(dR_{xy}/dB)|_{Bc}$ of the 4-3 transition with different excitation currents at the base temperature 1 mK.